\documentclass{aastex}          
\usepackage{spr-astr-addons}    

\begin{document}
%
\title{Morphological Analysis on the Coherence of kHz QPOs}

\shorttitle{Morphological Analysis on the Coherence of kHz QPOs}
\shortauthors{J. Wang et al.}

\author{J. Wang}
\author{H. K. Chang}
\affil{Institute of Astronomy, National Tsing Hua University, Hsinchu, 30013, Taiwan}
\email{joanwangj@126.com}
\and
\author{C. M. Zhang}
\affil{National Astronomical Observatories, Chinese Academy of Sciences, Beijing, 100012, China}
\and
\author{D. H. Wang\altaffilmark{1}}
\author{L. Chen}
\affil{Astronomy Department, Beijing Normal University, Beijing, 100875, China}
\and
\author{J. L. Qu}
\author{L. M. Song}
\affil{Institute of High Energy Physics, Chinese Academy of Sciences, Beijing, 100049, China}

\altaffiltext{1}{National Astronomical Observatories, Chinese Academy of Sciences, Beijing, 100012, China}

\begin{abstract}
We take the recently published data of twin
kHz quasi-period oscillations (QPOs) in neutron star (NS) low-mass
X-ray binaries (LMXBs) as the samples, and investigate the
morphology of the samples, which focuses  on the quality factor,
peak frequency of kHz QPOs,  and try to infer their physical
mechanism. We notice that: (1) The quality factors of upper kHz
QPOs are low ($2 \sim 20$ in general) and increase with the kHz QPO
peak  frequencies for both Z and Atoll sources. (2) The
distribution of  quality factor versus  frequency for the lower kHz
QPOs are quite different  between Z and Atoll sources.  For most Z
source samples, the quality factors of lower kHz QPOs are low
(usually lower than 15) and rise steadily with the   peak
frequencies except for Sco X-1, which  drop   abruptly at the
frequency of about $750$ Hz. While for most Atoll
sources, the quality factors of lower kHz QPOs are very high (from
2 to 200) and usually have a rising part, a maximum and an
abrupt drop. (3) There are three Atoll sources (4U 1728-34, 4U
1636-53 and 4U 1608-52) of  displaying  very high quality factors
for lower kHz QPOs. These three sources have been detected with the spin frequencies and sidebands,
in which the source with
higher spin frequency presents higher quality factor of lower kHz
QPOs  and lower difference between sideband frequency and lower kHz
QPO  frequency.
\end{abstract}


\section{Introduction}\label{s:1}

The kilohertz quasi-periodic oscillations in neutron star X-ray binaries were discovered
just two months after the launch of Rossi X-ray Timing Exploer
(RXTE) in Sco X-1 \citep{van der Klis96a,van der Klis96b,van der Klis96c} and 4U 1728-34
\citep{Strohmayer96a,Strohmayer96b,Strohmayer96c}. The former is a bright Z source,
while the latter is an Atoll source. In general, Z sources
produce the Z-shaped paths in CCDs or hardness-intensity diagram
(HID) on timescales of a few hours to days, with three branches,
from hard to soft: horizontal branch (HB), normal branch (NB), and
flaring branch (FB). Z sources are objects with high luminosity
($0.5-1.5 L_{Edd}$)\citep{van der Klis06} and mildly
high magnetic field \citep{Miller98,Campana00,Zhang07}.
The Atoll sources, with low luminosity ($\sim 0.001-0.2
L_{Edd}$)\citep{van der Klis06} and inferred weaker
magnetic field \citep{Miller98,Campana00,Zhang07}, show a hard, low-luminosity, fuzzy island state (IS) and a soft,
high-luminosity "banana" shaped structure on a timescale of
weeks \citep{Hasinger89,Hasinger90,van der Klis00,van der Klis06}. They trace a U-shaped or a C-shaped track as the
sources spectrum evolves between the island and the banana
\citep{Gladstone07}.

These kHz QPOs in LMXBs are peaks with some width in their power
density spectra (PDS), and their profiles can be described by
the Lorentzian Function $P_{\nu}\propto A_{0}
w/[(\nu-\nu_0)^2+(w/2)^2]$ \citep{van der Klis00,van der Klis06}. Here, $\nu_0$
is the peak frequency (or centroid frequency), $w$ is the full width
at half-maximum (FWHM), and $A_0$ is the amplitude of this signal.
The ratio of these two quantities is the quality factor, i.e., $Q
\equiv \frac{\nu_0}{w}$. Therefore, the kHz QPOs can be
characterized by three characteristic quantities, i.e., $\nu_0$,
quality factor (Q) and the fractional root-mean-squared (rms, which
represents a measure of the signal strength and is proportional to
the square root of the peak power contribution to the PDS). The
quality factor characterizes the coherence of a QPO signal, and its
value is related to the lifetime of the signal.

Most kHz QPO signals usually occur in twins. The centroid
frequencies (or peak frequency, i.e. upper $\nu_2$ and lower $\nu_1$
frequency) changes with accretion rate. Each peak has a corresponding quality factor
(i.e. upper $Q_2$ and lower $Q_1$). In the past few
years, many works have been dedicated to the investigation for
quality factor as a function of peak frequency of kHz QPOs
\citep{Barret05a,Barret05b,Barret05c,Barret06,Barret07,Barret08,
Boutelier09,Boutelier10}. Using RXTE
data, \citep{Barret05a} studied 4U 1608-52 and revealed a
positive correlation between $\nu_1$ and $Q_1$, up to a maximum of
about $Q \sim 200$. Motivated by this idea, \citep{Barret05b,Barret06}
studied, in a systematic way, the QPO properties of 4U
1636-53 and the dependency of $Q$ on $\nu$. It is shown that $Q_1$
and $Q_2$ of 4U 1636-53 follow different tracks in a $Q-\nu$ plot,
i.e. $Q_1$ increases with $\nu_1$ up to $850 Hz$ ($Q \sim 200$) and
drops precipitously to the highest detected frequencies $\sim 920
Hz$ ($Q \sim 50$), while $Q_2$ increases steadily all the way to the
highest detectable QPO frequency. Moreover, $Q_1$ is higher than
$Q_2$ \citep{Barret05b,Barret05c,Barret06}. The $Q - \nu$ distribution in the sources
4U 1735-44, 4U 1728-34 have the similar trend as 4U 1636-53
\citep{Barret06,Boutelier09,Mendez06,Torok09}.
Besides, \cite{Mendez06}
studied the relations between maximum amplitude and coherence of kHz
QPOs for a dozen of NS LMXBs.

Although most attentions for the coherence of kHz QPOs have been
focussed on the individual source, there are no work to perform
contrastive investigation for the properties between different types
of sources, e.g. Atoll and Z sources. Triggered by this background, in this paper, we study
the recently published data of $Q$ for twelve sources in terms of
Atoll and Z types, i.e. seven Atoll sources (4U 1608-52, 4U 1636-53,
4U 1728-34, 4U 0614+09, Aql X-1, 4U 1820-30 and 4U 1735-44) and five
Z sources (Sco X-1, Cyg X-2, GX 17+2, GX 5-1 and GX 340+0). We
benefit from the existing studies and use the published data from
the collection by \citep{Mendez06}. Besides, we also consider
the source XTE J1701-462, a peculiar source exhibiting typical Z
behaviors in CCDs \citep{Remillard06,Homan06a,Homan06b,Homan07} and evolving into an Atoll source at the end of its
outburst \citep{Homan10}. In section 2, we make a contrastive
investigation for the quality factor as a function of kHz QPO peak frequency. The
implications are discussed in section 3. Section 4 contains the
summary.

\section{Contrastive Analysis for the Coherence of KHz QPOs}\label{s:2}

In this section, we study the recently published data of quality
factor as a function of peak frequency for seven Atoll and five Z
sources, in order to find out the differences between Z and Atoll
sources. At first, we put all the sources together and study
the relations for $Q_1-\nu_1$ and $Q_2-\nu_2$. Secondly, we
investigate their distinctive properties for
Z and Atoll sources, respctively. All the data are obtained by Proportional
Counting Array (PCA) on board the RXTE (see M\'{e}ndez 2006 and
references therein for the data selection).
For the data, no filtering on the raw data is performed, i.e.
all photons are used and only time intervals containing X-ray bursts
are removed \citep{Barret05a,Barret05b,Barret05c,Barret06,Barret07}.

\subsection{The $Q-\nu$ Relations for Upper and Lower kHz QPOs}\label{ss:1}

\begin{figure*}
\includegraphics[width=8.35cm]{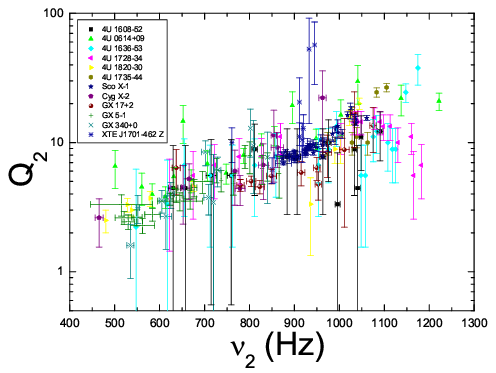}
\includegraphics[width=8.35cm]{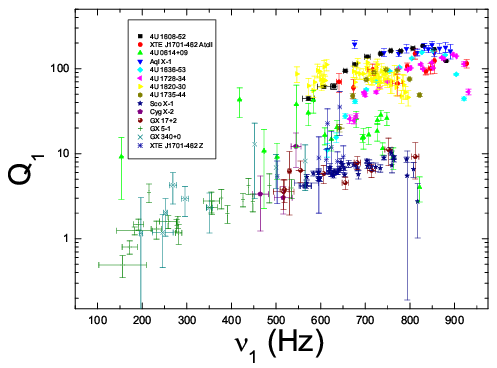}
\caption{Plots for $Q-\nu$ relations of kHz QPOs. The left panel is
the $Q-\nu$ relations for upper kHz QPOs, and the right one is that
for lower kHz QPOs. The meanings of different symbols are listed in
the diagrams.}
\label{Fig:total}
\end{figure*}

We put the twelve sources together and investigate the evolution of
Q as a function of $\nu$ (see Fig. \ref{Fig:total}). From Fig.
\ref{Fig:total}, we find that the values of $Q_2$ (in general, $Q_2
\sim 2-20$) are lower than those of $Q_1$ (up to 200) as a whole.
The Q changes with $\nu$ very differently for upper and lower
kHz QPOs, and the ranges of Q are also different.

For the upper kHz QPOs (see the left panel of Fig. \ref{Fig:total}),
most data points of $Q_2$ locate in the range with low values, i.e.
$Q_2 \sim 2-20$ and $\nu_2 \sim 450-1200$ Hz. It can be seen
that most sources follow the steadily rising track in the
$Q_2-\nu_2$ plot. At the frequency of $\nu_2 \sim 1050 Hz$, a
transition presents for 4U 1728-34, i.e. $Q_2$ begins to drop with
$\nu_2$. The $Q_2$ of XTE 1701-462 in Z phase present very large
errors, clustering in a very narrow range of frequency.

As far as the profiles of lower kHz QPOs, It displays two different
trajectories in the $Q_1-\nu_1$ plot (see the right panel of Fig.
\ref{Fig:total}). The data points of Z sources follow the steadily
rising tendency with low values of $Q_1$ ($Q_1 \sim 2-20$), and
$\nu_1$ covers a wide range of frequency ($\nu_1 \sim 150-825$ Hz).
However, most $Q_1$ for Atoll sources increase with $\nu_1$ to a
maximum and abruptly drop. The $Q_1$ can be very high and vary from 2
to 200. The centroid frequencies for these sources cover a
relatively narrow range $\sim 550-950$ Hz. Moreover, despite of the
similar tendency, different sources show their own "substructures"
in different regions in $Q_1-\nu_1$ diagram, which we will discuss
in detail in the following.

On the whole, the Q-factors of lower kHz QPOs for Atoll sources are
about 10 times higher than that for Z sources for the same frequency.

\subsection{The $Q-\nu$ Correlations for Z Sources}\label{ss:2}

\begin{figure*}
\includegraphics[width=8.3cm]{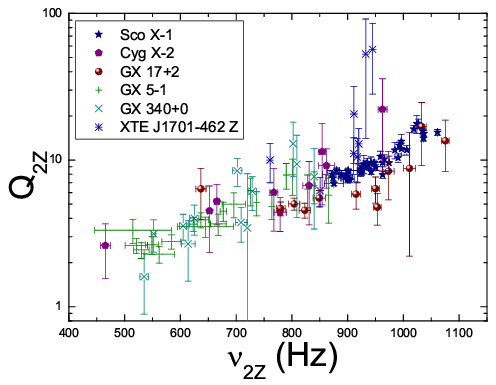}
\includegraphics[width=8.55cm]{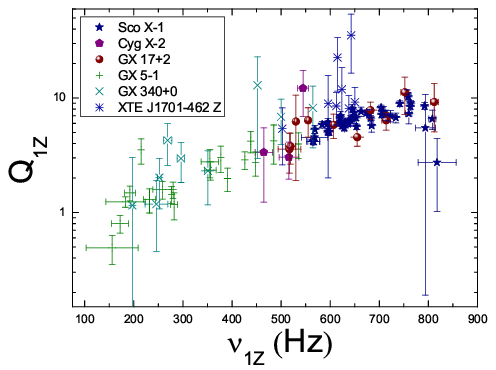}
\caption{Plots for $Q-\nu$ relations for Z sources. The left panel
is for upper kHz QPOs, and the right one is for lower kHz QPOs. The
meaning of different symbols are listed in the diagrams.}
\label{Fig:Z}
\end{figure*}

Two plots in Fig. \ref{Fig:Z} present the regular and steady
tendency, formed by five Z sources and the Z phase of XTE
1701-462. It can be found that the $Q-\nu$ tracks of Z sources are
similar between upper and lower kHz QPOs. For a close inspection, we
find that different sources lie in different regions in the $Q-\nu$
plots, and it can be divided into two regions by the range of
frequency for both upper and lower kHz QPOs. One contains GX 5-1 and
GX 340+0, with low Q and $\nu$. The other includes Sco X-1 and GX
17+2, which presents relatively high Q and $\nu$.

The detected $\nu_1$ and $\nu_2$ for GX 5-1 range from
156$\pm$23 Hz to 627$^{+25}_{-13}$ Hz and from 478$\pm$15 to
866$\pm$23 Hz, respectively \citep{Jonker02}. For GX 340+0,
$\nu_1$ and $\nu_2$ increase from 197$^{+26}_{-70}$ Hz to
565$^{+9}_{-14}$ Hz and from 535$^{+85}_{-48}$ to 840$\pm$21 Hz,
respectively \citep{Jonker00a,Jonker00b}. These two sources form the lower
region in both $Q_1-\nu_1$ and $Q_2-\nu_2$ plots.
The real maximum may larger than the largest measured value , in fact,
the maximum can be at most $\sim 20 \%$ higher
than the values we used \citep{Mendez06}, so it seems that the
real maxima are too different from the largest values we used in Fig.
\ref{Fig:Z}.  The same
is true for $Q_{max}$ and $\nu_{max}$ of other sources. The $\nu_2$
and $\nu_1$ in the other region (including Sco X-1 and GX 17+2) are
$\nu_2 \sim 850 - 1150$ Hz and $\nu_1 \sim 500-800$ Hz \citep{Homan02,van der Klis97}.
In addition, Cyg X-2 covers
almost the whole range of $Q_2$ and $\nu_2$ that the others cover,
but a very narrow range for both $Q_1$ and $\nu_1$. Because
there are just three measurements of $Q_1$ for this source, it is
difficult to infer any implication.

XTE J1701-462 presents the link between "Z-track", "$\nu$-track" and
Atoll behavior in its CCD and HID \citep{Homan06a,Homan06b,Homan07,Lin09,
Homan10}. Both upper and lower QPOs were detected in
Z phase \citep{Sanna10}. The Z phase of XTE
J1701-462 presents quite large errors for both $Q_1$ and $Q_2$, and
the frequencies cover the narrow range. The $Q_2$ are higher than
$Q_1$ for this source, which is different with that of the other Z
sources. We think the transition between Atoll and Z state of XTE J1701-462 may link to the accretion rate. When the companion is near periastron, the neutron star may have higher accretion rate, its magnetosphere will compress and show a higher magnetic field, then the system shows Z state. On the contrary, when the companion is near apastron, the neutron star may have lower accretion rate, the magnetosphere will expand and show a lower magnetic field, then the system shows Atoll state. But it needs more observational and theoretical investigation.

It is noticed that most data points follow an increasing trend in
$Q_1-\nu_1$ diagram (see the right panel of Fig. \ref{Fig:Z}), but
an abrupt drop occurs for Sco X-1 at about $750$ Hz.
Zhang et al. investigated the correlation between upper and lower kHz QPOs in a statistical way \citep{Zhang06}. They found the power law correlation can fit the data much better than other correlations, e.g. linear relation and constant relation. At the same time, they also found that the power law index shows a turn-over frequency at about $\nu_2=840 Hz$ in $\nu_2~vs.~\nu_1$ plot of Z source, which is analogical with the abrupt drop in $Q_1~vs.~\nu_1$ plot of Sco X-1. But, we are not sure whether it results from the physical process or from the data itself. If it is a physical reason, it
implies that the boundary of the accretion disk and magnetosphere, at where the kHz QPOs are
assumed to  emit this particular frequency,
may occur a physical transition. From the Alfv\'en wave model for
kHz QPOs \citep{Zhang04}, the disk radius of emitting this particular
frequency  is  about 20 km, or 5 km away from the stellar
surface for the star parameters of 15 km and one solar mass \citep{Zhang10}.

\subsection{The $Q-\nu$ Correlations for Atoll Sources}\label{ss:3}

\begin{figure*}
\includegraphics[width=8.4cm]{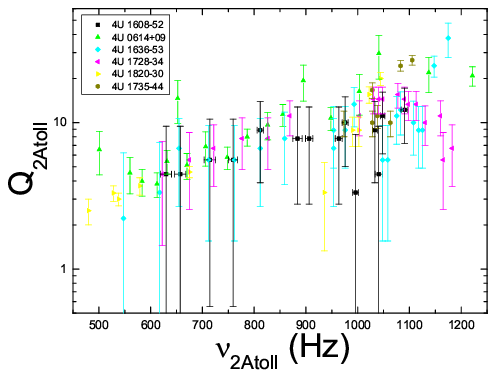}%
\includegraphics[width=8.4cm]{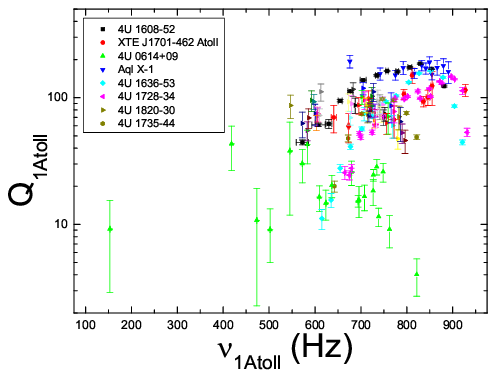}
\caption{The same meaning as Fig. \ref{Fig:Z}, but for Atoll sources.}
\label{Fig:Atoll}
\end{figure*}

From the left panel of Fig. \ref{Fig:Atoll}, it is seen that most $Q_2$
increases with $\nu_2$ steadily, only the source 4U
1728-34 shows an explicit drop at about $\nu_2 \sim 1090.4$ Hz with a maximum of $Q_2 =
14.4\pm3$. The Kepler orbital radius corresponding to 1090.4 Hz is about $\sim17km$ for a NS of 1.4 solar mass. Thus, the position of turn-over frequency is close to the innermost boundary of the accretion disk and it may reflect the corresponding physical process there. The innermost stable circular orbit (ISCO) of a NS with mass of 1.4 solar mass is $\sim12.6 km$, which is the same order as the theoretical radius of NS. We do not know the actual NS mass of 4U 1728-34, so we are not sure whether the ISCO is larger than the star radius. We think that, no matter in which case, the accretion matter may drop because of entering the ISCO boundary or impact onto the surface of the star, and then the system may show an abrupt drop of Q factor.
However, the right panel of Fig. \ref{Fig:Atoll} presents a very
different scenario. Almost all sources display rising, maximum and
dropping tendency in the $Q_1-\nu_1$ diagram. Five sources (4U
1608-52, 4U 1636-53, 4U 1728-34, 4U 1820-30 and 4U 1735-44) present
obvious dropping tendency after the maximum $Q_1$. The maximums are
different for each source. 4U 1608-52, 4U 1636-53, 4U 1728-34
present higher $Q_1$ (The detected maxima are 247.0$\pm$16.0,
248.0$\pm$18.0, and 188.0$\pm$18.0, respectively) with wider $\nu_1$
ranges. The narrowest $\nu_1$ range of 4U 1735-44 is from 642 to 821
Hz. There is also a gap of $\nu_1$ (between 613 and 673 Hz) in 4U
1820-30. The $Q_1$ of Aql X-1 are high and cover a very narrow
range ($151\pm13.2 - 193.8\pm22.2$), and so does that of the peak
frequency ($696.1\pm0.15 - 890.7\pm0.37$ Hz). It seems that the kHz
QPO frequencies just are detected at the inflexion of $Q_1-\nu_1$
track. The $Q_1$ for 4U 0614+09 are relatively low and irregular
with large errors, and the maximum is $43.12\pm16.45$ at
$\nu_1=418.3\pm1.8$ Hz. The points which are higher than $Q_1=30$
presents very large error bars, and most of the $Q_1$ are lower than
30. For the Atoll phase of XTE J1701-462, we use the data from the
paper by \citep{Sanna10} for our plot. It is found that the
maximum $Q_1$ is $150.3\pm20.9$ at $811.5\pm6.6$ Hz, and above this
frequency the $Q_1$ values begin to drop.

\section{Discussions}\label{s:3}

\subsection{Relations between $Q-\nu$ Tracks and CCDs Tracks for Z Sources}\label{ss:1}

In spite of some similar respects \citep{Stella86}, the tracks in CCDs
and HIDs for Z sources, by secular timing inspection, can be divided
into two types according to qualitative differences \citep{Kuulkers94}, i.e. Cyg-like sources (Cyg X-2, GX 5-1 and GX 340+0)
and Sco-like sources (Sco X-1 and                                                                                                         GX 17+2)\citep{Hasinger89,Hasinger90}. Recently, it is
claimed that the Cyg-like sources follow a "Z-track" in CCDs
which have a long horizontal branch and form a "Z" profile, while
the Sco-like sources present a "$\nu$-track" which have a short
horizontal branch and form a "$\nu$" profile \citep{Homan07}. According to the recent data obtained with PCA on
board RXTE, the five typical sources (Cyg X-2, GX 5-1, GX 340+0, Sco
X-1 and GX 17+2) can be divided into two subclasses. Cyg X-2, GX
5-1, and GX 340+0 present "Z-track" \citep{Jonker00a,Jonker00b,Jonker02,Wijnands98}. Besides, GX 17+2 and Sco X-1
exhibit "$\nu$-track" with not well defined HB/NB vertex, since the
HB is almost a continuation of the NB \citep{Homan02}. The kHz
QPOs are detected on the vertex of HB/NB and the upper NB. It is
claimed that the properties of kHz QPOs closely related to the
position of the sources on the Z track traced out in CCDs, and the
frequencies of kHz QPOs increase from the left of HB to NB/FB vertex
\citep{Wijnands97}.

It seems that the division between Sco-like and Cyg-like
sources can be in terms of frequency for kHz QPOs. The Cyg-like
sources present much longer HB, and it is found that the kHz QPOs in
this class of sources are observed at lower frequencies compared
with the other two Z sources \citep{Jonker00a,Jonker00b,Jonker02,Wijnands98}.
In addition, a transition from Cyg-like phase to Sco-like
phase occurs in Cyg X-2. This property may be consistent with its
wide $\nu_2$ range and the kHz QPOs at relatively high frequencies.
However, the $Q_1$ and $\nu_1$ lie in a very narrow region, but with
large error. The physical reason why the other sources don't cover
the whole range and Cyg X-2 does will be investigated in our future
work. Sco X-1 is a very bright Z source, in which the kHz QPOs
can be detected all the way onto the FB, corresponding to the high
frequencies that are observed. Besides, the high luminosity corresponds to the high accretion rate and it may have some possible reasons: One possibility is that the accretion disk is thickened by the radiation pressure during the Z stage. As a result, the magnetosphere of neutron star expands, while both the inner disk radius and the mass accretion rate increase. The signal may be sucked by the thickened disk and the system shows a drop of Q factor. Another possibility is that the innermost boundary of accretion disk is close to the NS surface. Consequently, the radius of magnetosphere is relatively small, and the frequency of kHz QPO can be up to a high value. The region between innermost disk and the NS surface become narrow. So the $Q_1$ can be up to the relatively high values. When the innermost disk is more close to NS surface, the radiation is sucked by some material accreted by the NS. As a result, the $Q_1$ begins to drop abruptly with $\nu_1$.

\begin{table*}
\small
\caption{The fitting results for $Q_{1max}$ of three Atoll sources.}
\label{tab:Qmax}
\begin{tabular}{ccccc}
\tableline
Source &$Q_{1max}$ &error(Q) &$\nu_1$ &error($\nu_1$)\\
\tableline
4U 1608-52 &182.9 &3.2 &824.03 &3.02\\
4U 1636-53 &168.15 &4.2 &833.9 &2.28\\
4U 1728-34 &143.61 &2.9 &872.57& 3.2\\
\tableline
\end{tabular}
\end{table*}

\begin{table*}
\small
\caption{The fitting results for spin frequency versus $Q_{1max}$.}
\label{tab:spin}
\begin{tabular}{ccccc}
\tableline
Relation &a &b &c & d \\
\tableline\\
$y=a^{bx-c}+d$ &1.0216 &0.98602 &438.85 &146.22\\
\tableline
\end{tabular}
\end{table*}

\begin{table*}
\small
\caption{The fitting results for the difference between $\nu_1$ and
sideband frequency versus $Q_{1max}$.}
\label{tab:siddif}
\begin{tabular}{ccccc}
\tableline
Relation &a &b &c &d  \\
\tableline
$y=a^{bx-c}+d$ &0.89119& 0.44606& 62.50412 &96.6845 \\
\tableline
\end{tabular}
\end{table*}

\begin{figure*}
\includegraphics[width=8.5cm]{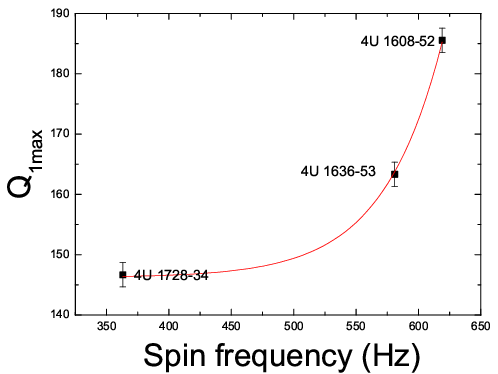}
\includegraphics[width=8.5cm]{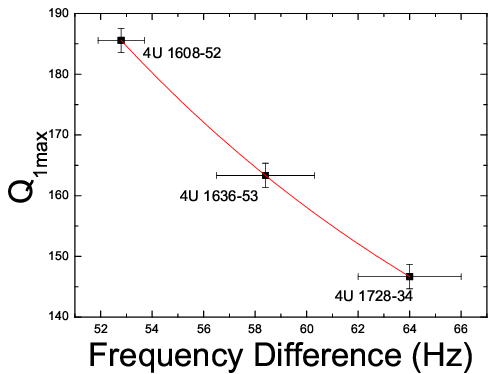}
\caption{The fittings for the correlations between spin frequency /
frequency difference and $Q_{1max}$. The left panel is for the spin
frequency versus $Q_{1max}$. The right one is for the difference
between $\nu_1$ and sideband frequency versus $Q_{1max}$.}
\label{Fig:fit}
\end{figure*}

\subsection{{\bf The correlation among ${\bf Q-\nu}$ track, spin frequency and sideband frequency}}\label{ss:2}

The peculiar trends of $Q_1$ as the function of $\nu$ for Atoll
sources imply the special physics in the inner disk region of these
class of sources. The high values of $Q_1$ indicate the small range
of frequency drift for lower kHz QPOs. It is claimed that the drop
of $Q_1$ is a hint of the innermost stable boundary of accretion
disk \citep{Miller98,Barret05c}.
The maximum of $Q_1$ is detected at a mildly higher frequency than
that of Z sources which is consistent with the small radius of
innermost disk. However, the Atoll sources present low luminosity
\citep{van der Klis06} and low accretion rate. Accordingly, the
magnetic field should be mildly weaker than that of Z sources,
allowing for the accretion disk extending to be close to the NS
surface.

From the second part of section \ref{s:2}, we know that three sources
(4U 1608-52, 4U 1636-53 and 4U 1728-34) exhibit very high $Q_1$
values. Moreover, the spin frequencies have been detected for these
sources, i.e. $619 Hz$ \citep{Hartman03}, $581 Hz$ \citep{Strohmayer98,Wijnands97,Zhang97},
$363 Hz$
\citep{Strohmayer96c}, respectively. In addition, all the three
sources display the sideband in their PDS \citep{Jonker00a,Jonker00b}. The differences in frequencies between the
sidebands and the lower kHz QPOs are, $52.8\pm0.9 Hz$ (4U 1608-52),
$58.4\pm1.9 Hz$ (4U 1636-53) and $64\pm2 Hz$ (4U 1728-34)\citep{Jonker00a,Jonker00b}. Maybe the high values
of $Q_1$ of the three sources allow for the detection of sidebands,
while the sidebands may be engulfed by the boarder peak in other
sources.

In order to investigate the relations between spin frequency /
sideband frequency and the maximum $Q_1$, we fit the function GCAS
to the data of these sources and find out the maximum $Q_1$ for
every sources (see Table \ref{tab:Qmax}). Then we plot the spin
frequency versus $Q_{1max}$ and the difference between sideband
frequency and the lower frequency versus $Q_{1max}$. We also
fit an exponential relation to them (see Fig. \ref{Fig:fit}). The
fitting results are listed in Table \ref{tab:spin} and Table
\ref{tab:siddif}. We notice that the source with higher maximum
$Q_1$ present higher spin frequency (see the left panel of Fig.
\ref{Fig:fit}), while the difference between $\nu_1$ and the
sideband frequency is low (see the right panel of Fig.
\ref{Fig:fit}).

The high maximum of $Q_1$ implies that the emission site of lower
kHz QPOs is more close to the NS surface. Besides, the high spin
frequency is consistent with a small corotation radius \citep{Zhang07}. With small
corotation radius, the drift range of frequency between this radius
and the NS surface is narrow, which corresponds to the high value of
$Q_1$. Accordingly, this give us the hint that the lower kHz QPOs
may relate to the corotation radius. The source with high spin
frequency and high maximum value of $Q_1$ presents the small
difference between sideband frequency and $\nu_1$. So we claim that
the emissions of lower kHz QPOs and the sideband are spin mediated.

\subsection{Revelation for The Nature of kHz QPOs}\label{ss:3}

The upper frequency is the same order as the dynamical time-scales
of the innermost region of the accretion flow around the stellar
mass compact objects (van der Klis 2006, 2008). It is considered
that $\nu_2$ is the innermost orbital frequency of accretion flow
both for Z and Atoll sources. Due to some instabilities resulting
from the changes of accretion rate, magnetic pressure and others
\citep{Romanova07,Rastatter99,Kulkarni08}, the boundary of inner
disk is changing during the accretion process. As a consequence,
every peak of kHz QPO signal presents drift around the centroid
frequency, contributing to the upper quality factor $Q_2$ \citep{Wang11}.

As far as the nature of lower kHz QPOs, two very different
evolutionary scenarios of $Q_1$ as the function of $\nu_1$ (see the
right panels of Fig. \ref{Fig:Z} and Fig. \ref{Fig:Atoll} for a
detail) imply distinct physics in the inner disk for Z and Atoll
sources. High accretion rate leads to strong instabilities
which are responsible for the large frequency drift. So the values
of $Q_1$ for Z sources are low. For the Atoll sources, the accretion
flow with low accretion rate and the relatively stable scenario in
the inner region of disk allows for the disk extending all the way
to near the NS surface, which accounts for the high values of $Q_1$.
We expect to find out the hints for mechanism of lower kHz QPOs from
the nature of Z and Atoll sources in our future work.

\section{Summary}\label{s:4}

More and more data about the coherence of kHz QPOs in NS LMXBs are
detected, the $Q-\nu$ relations and their implications have been the
attractive issues. In this paper, we investigate the recently
published data of quality factors for thirteen sources
\citep{Mendez06,Sanna10}, i.e. seven Atoll sources
(4U 1608-52, 4U 1636-53, 4U 1728-34, 4U 0614+09, Aql X-1, 4U 1820-30
and 4U 1735-44), five Z sources (Sco X-1, Cyg X-2, GX 17+2, GX 5-1
and GX 340+0) and XTE 1701-462 which presents both Z and Atoll
behaviors. The main conclusions are listed below.

(1). The $Q_2$ values are low ($Q_2 \sim 2-20$) for both Z and Atoll
sources. The $Q_2-\nu_2$ tracks increase steadily, in general. The
values of $Q_1$ are are low ($Q_1 \sim 2-20$) for Z sources and
increase with frequencies(except for Sco X-1). But the
values of $Q_1$ for Atoll
sources are very high (up to 200), and increases with frequencies up
to a maximum then abruptly drops.

(2). Though the $Q-\nu$ distribution of Z sources form a continues relation for both upper and lower
kHz QPOs, they can be divided into two regions, according to the ranges of
$\nu$. One contains Sco X-1 and GX 17+2, which present high Q-values
and high centroid frequencies. The other is formed by GX 5-1 and GX
340+0, with low Q-values and centroid frequencies. However, Cyg X-2
extends almost to the whole range of $Q_2$ and $\nu_2$ that the
others cover. The different ranges of frequency may give us
useful information about the nature of Z and Atoll source.

(3). The $\nu_2$ for Atoll sources cover a boarder range and $Q_2$
values are low $2-20$. Almost all $Q_1-\nu_1$ tracks for Atoll
sources present rising part, maximum and then the abrupt drop.

(4). The spin period and sidebands were detected in three sources
(4U 1608-52, 4U 1636-53 and 4U 1728-34) which present very high
$Q_1$. The source with higher spin frequency presents higher $Q_1$
values, and its difference between lower $\nu_1$ and sideband
frequency is low.

(5). The $Q_1$ values are the same order as $Q_2$ for 4U 0614+09.
The lower frequencies for Aql X-1 just were detected in a very
narrow range, and the $Q_1$ values are high. XTE 1701-462 presents
high errors for $Q_2$ and $Q_1$.

(6). The emission of lower kHz QPOs and sideband frequency may
be correlated to the NS spin.

\acknowledgments
We acknowledge M. Mendez and D. Barret for providing the data. It is also a pleasure to thank M.
A. Abramowicz, G. Q. Ding, J. Hor$\acute{a}$k and W.
Klu$\acute{z}$niak for helpful discussions. This work is supported
by National Basic Research Program of China (2009CB824800, 2012CB821800),
the National Natural Science Foundation of China (NSFC 10773017, NSFC 10773034, NSFC 10778716, NSFC 11173024),
NSC 99-2112-M-007-017-MY3 and the Fundamental Research Funds for the Central Universities.

\end{document}